\documentclass[12pt]{article}

\usepackage[height=8.85in,width=6.45in]{geometry}
\usepackage{amsmath,amssymb}
\usepackage{hyperref}

\def\tr{\mathop{\mathrm{tr}}\nolimits}
\def\Tr{\mathop{\mathrm{Tr}}\nolimits}

\usepackage{times}
\usepackage{courier}
\usepackage{mathtools}
\numberwithin{equation}{section}

\let\bar\overline

\def\Nequals#1{$\mathcal{N}{=}\,#1$}

\newcommand{\bC}{\mathbb{C}}
\newcommand{\bH}{\mathbb{H}}

\newcommand{\bZ}{\mathbb{Z}}

\newcommand{\SU}{\mathrm{SU}}
\newcommand{\UU}{\mathrm{U}}
\newcommand{\OO}{\mathrm{O}}
\newcommand{\Sp}{\mathrm{Sp}}
\newcommand{\SO}{\mathrm{SO}}

\newcommand{\fsu}{\mathfrak{su}}
\newcommand{\fg}{\mathfrak{g}}

\newcommand{\be}{\begin{equation}}
\newcommand{\ee}{\end{equation}}
\newcommand{\bea}{\begin{eqnarray}}
\newcommand{\eea}{\end{eqnarray}}

\begin{document}

\begin{titlepage}

\begin{flushright}
IPMU-17-0032
\end{flushright}

\vskip 4cm

\begin{center}

{\Large\bfseries Anomaly matching on the Higgs branch}

\vskip 1cm
Hiroyuki Shimizu, Yuji Tachikawa and Gabi Zafrir
\vskip 1cm

\begin{tabular}{ll}
  & Kavli Institute for the Physics and Mathematics of the Universe, \\
& University of Tokyo,  Kashiwa, Chiba 277-8583, Japan\\
\end{tabular}

\vskip 1cm

\end{center}

\noindent
We point out that we can almost always determine by the anomaly matching the full anomaly polynomial of a supersymmetric theory in 2d, 4d or 6d  
if we assume that its Higgs branch is the one-instanton moduli space of some group $G$.
This method not only provides by far the simplest method to compute the central charges of known theories of this class, e.g.~4d $E_{6,7,8}$ theories of Minahan and Nemeschansky or the 6d E-string theory, but also gives us new pieces of information about unknown theories of this class.

\end{titlepage}

\section{Introduction}

Instantons of classical groups can be described in terms of the ADHM construction \cite{Atiyah:1978ri}, which can in turn be realized as the Higgs branch of supersymmetric gauge theories \cite{Witten:1995gx,Douglas:1996sw}. 
These gauge theories arise as the worldvolume theories on perturbative $p$-branes probing  perturbative $(p+4)$-branes, and the motion into the Higgs branch corresponds to the process where $p$-branes get absorbed as instantons of the gauge fields on $(p+4)$-branes.

In string/M/F theory, there are also non-perturbative branes that host exceptional gauge groups, and if we probe them by lower-dimensional branes, we get supersymmetric theories whose Higgs branch equals to the instanton moduli spaces of exceptional groups. 
Among them we can count the 4d theories of Minahan and Nemeschansky \cite{Minahan:1996fg,Minahan:1996cj} for $E_{6,7,8}$ instantons and the 6d E-string theory \cite{Ganor:1996mu,Seiberg:1996vs}.

The theories obtained this way do not usually have any conventional Lagrangian descriptions, and were therefore rather difficult to study. 
Even their anomaly polynomials, or equivalently the conformal central charges assuming that they become superconformal in the infrared, needed to be computed first with stringy techniques \cite{Aharony:2007dj,Ohmori:2014pca} and then with rather lengthy field theoretical arguments on the Coulomb branch in 4d or on the tensor branch in 6d \cite{Shapere:2008zf,Ohmori:2014kda,Intriligator:2014eaa}.

In this paper, we point out that the anomaly matching on the Higgs branch almost always allows us to determine the full anomaly polynomial, when the theory is 6d \Nequals{(1,0)}, 4d \Nequals2, or 2d \Nequals{(4,0)}, and when the Higgs branch is assumed to be the one-instanton moduli space of some group $G$. 
This is because on the generic point of the Higgs branch  the theory becomes free and the unbroken symmetry still knows the $\SU(2)_R$ symmetry at the origin.

This method provides the simplest way to compute the anomaly polynomials of 4d theories of Minahan and Nemeschansky and the 6d E-string theory. 
But more importantly, this method gives us new pieces of information about a theory whose Higgs branch is the one-instanton moduli space of the group $G$, even when no string/M/F theory construction is known. 
For example, in \cite{Beem:2013sza}, the conformal bootstrap method was used to determine the conformal central charges of the 4d theory whose Higgs branch is the one-instanton moduli space of $G_2$ or $F_4$. 
Our method reproduces the values they obtained, and not only that, we find a strong indication that the $F_4$ theory does not exist because of a field theoretical inconsistency. 
Similarly, we will see that there cannot be any 6d $E_{6,7}$ theory.

The rest of the paper is organized as follows. 
In Section 2, we describe in more detail how the anomaly matching on the Higgs branch works if the Higgs branch is the one-instanton moduli space of some group $G$. In Section 3, we summarize the results which we obtained in this paper. 
Then in Section 4, 5, 6, we study the 6d \Nequals{(1,0)} theories, the 4d \Nequals2 theories,  and the 2d \Nequals{(4,0)} theories in turn.
In Appendix A, we collect the formulas for characteristic classes used throughout in this paper.

\section{Basic idea}
We consider a theory with 6d \Nequals{(1,0)} or 4d \Nequals2 or 2d \Nequals{(4,0)} supersymmetry has a Higgs branch given by the one-instanton moduli space $M_G$ of a group $G$.

\paragraph{Geometric data:}
Let us first recall some basic information on $M_G$, whose detail can be found e.g.~in \cite{Gaiotto:2008nz} and the references therein.
The quaternionic dimension of $M_G$ is $h^\vee(G)-1$.
We note that for $G=\Sp(n)$, the one-instanton moduli space is simply $\bH^n/\bZ_2$, where $\bH$ is the space of quaternions.

Furthermore, the moduli space is smooth on a generic point,
and the symmetry $\SU(2)_R\times G$ acting on  $M_G$ 
is broken to $\SU(2)_D\times G'$, where  $\SU(2)_X \times G' \subset G$ is a particular subgroup described in more detail below and $\SU(2)_D$ is the diagonal subgroup of $\SU(2)_R$ and $\SU(2)_X$. 
The subgroup $\SU(2)_X$ is the $\SU(2)$ subgroup associated to the highest root of $G$ and $G'$ is its commutant within $G$. The tangent space of $M_G$ at a generic point transforms under $\SU(2)_X \times G'$ as a neutral hypermultiplet and a charged half-hypermultiplet in a representation $R$, with the rule \begin{equation}
\fg=\fg' \oplus \fsu(2) \oplus R.
\end{equation} Here $R$ is always of the form of the doublet of $\SU(2)_X$ tensored with a representation $R'$ of  $G'$.
The subgroup $G'$ and  the representation $R'$ are given in the table~\ref{basicdata}.

\begin{table}
\[
\begin{array}{c||c|c|c|l}
G & h^\vee & G' & R' &\text{short comment}\\
\hline
\SU(n) & n& \UU(1)_F\times \SU(n-2)   &  (\mathbf{n-2})_{-n} \oplus (\overline{\mathbf{n-2}})_{+n}
&   \\
\SO(n) & n-2 & \SU(2)_F\times \SO(n-4)  & \mathbf{2}_F \otimes (\mathbf{n-4})\\
\Sp(n) & n+1 & \Sp(n-1) & \mathbf{2n-2}\\
E_6 & 12 & \SU(6) & \mathbf{20} & \text{3-index antisym.}\\
E_7 & 18 & \SO(12) & \mathbf{32} & \text{chiral spinor.} \\
E_8 & 30 & E_7 & \mathbf{56} \\
F_4 & 9 & \Sp(3) & \mathbf{14}' & \text{3-index antisym.~traceless.}\\
G_2 & 4 & \SU(2)_F& \mathbf{4} & \text{3-index sym.} 
\end{array}
\]
\caption{The data. 
For $\SU(n)$, $\UU(1)_F$ is normalized so that $\mathbf{n}$ splits as $(\mathbf{n-2})_{-2}$ and $\mathbf{2}_{n-2}$. For $\SO(n)$, $n$ is assumed to be $\ge 5$.
\label{basicdata}}
\end{table}

\paragraph{Strategy of the matching:}
Now let us explain how the anomaly matching on the Higgs branch works.
At the origin, the theory has the symmetry $\SU(2)_R\times G$ where $\SU(2)_R$ is (part of) the R-symmetry.

On a generic point of the Higgs branch, 
we have a free theory whose unbroken symmetry is $\SU(2)_D \times G'$,
where $\SU(2)_D$ is the diagonal subgroup of $\SU(2)_R$ and $\SU(2)_X$.
The theory is a collection of $d_H=h^\vee-1$ hypermultiplets. One, identified with changing the vev, is neutral under the unbroken global symmetry. Considering that the scalars in a half-hyper are doublets of $SU(2)_R$, this hyper should just be a half-hyper in the $\bold{2}$ of $SU(2)_X$. Additionally we have the remaining $d_H - 1$ hypers which transform as a doublet of $SU(2)_X$ and in some representation $R'$ of $G'$ given in table \ref{basicdata}. Since $SU(2)_X$ and $SU(2)_R$ are identified to be $\SU(2)_D$, this amounts to just $d_H - 1$ free hypers in the representation $R'$.

The anomaly of $G$ of the original theory can be found from the anomaly of $G'$ of the free theory in the infrared, if $G'$ is nonempty. 
This in turn determines the contribution of $\SU(2)_X$ to the anomaly of $\SU(2)_D$,
which then fixes the anomaly of $\SU(2)_R$ of the original theory.
Even if $G'$ is empty, this still constrains the anomaly of $\SU(2)_R$ and $G$ of the original theory.
Along the way, we might find that the anomaly matching cannot be satisfied, in which case we conclude that such a theory cannot exist. 
There are cases where the anomaly polynomials can be arranged to match but the global anomaly fails to match.\footnote{The authors thank Kazuya Yonekura for the discussions on this point.}
We call this the global anomaly matching test.

Finally, since the one-instanton moduli space of $\Sp(n)$ is $\bH^n/\bZ_2$ as explained below, we should always be able to match the anomaly in this case by $n$ free hypermultiplets gauged by $\bZ_2$, or equivalently an $\mathrm{O}(1)-\Sp(n)$ bifundamental gauged by $\mathrm{O}(1)$.
This provides us a simple way to check the computations.

Now that the strategy has been explained, we move on to the details. 
We first summarize the results in the next section, and then look at the three cases in turn, in the order 6d, 4d and 2d.

\section{Summary of results}
In this section we summarize the results we obtain in each spacetime dimensions, postponing the computational details in the following sections.
We  assume that there are just free hypermultiplets on the generic point on the Higgs branch unless otherwise stated. 

\subsection{Six-dimensional theories}
First we consider 6d \Nequals{(1,0)} theories.
We find that the anomaly polynomials on the Higgs branch can consistently be matched for 
\begin{equation}
\SU(2), \quad \SU(3),\quad \Sp(n),\quad  E_{8}, \quad \text{and}\quad G_2. 
\end{equation}
\begin{itemize}
\item In the $\SU(2)$ case, we cannot completely determine the anomaly at the origin; we find a three-parameter family of solutions \eqref{6dsu2}. 
The result is consistent with one known example, which is just a free hypermultiplet gauged by $\bZ_2$.
\item In the $\SU(3)$ case, we can unambiguously determine the anomaly as in \eqref{6dsu3}. But we do not know any  example of 6d theories with this Higgs branch.  
\item The $\Sp(n)$ case reproduces the anomaly polynomial of $n$ free hypermultiplets gauged by $\bZ_2$.
\item The $E_8$ case reproduces the anomaly of the rank-$1$ E-string theory. 
\item The $G_2$ case does not pass the anomaly matching test of the global anomaly, as detailed in Sec.~\ref{sec:6dglobal}
\end{itemize}

\subsection{Four-dimensional  theories}

\begin{table}[h]
\[
\begin{array}{c||c|cc|cc}
G & k & n_v & n_h & a & c  \\
\hline
\SU(2)  & x+1 & x & x+1 & \frac{6x+1}{24} & \frac{3x+1}{12} \\
\SU(3) & 3 & 2 & 4 & \frac{7}{12} & \frac{2}{3} \\
\SO(8) & 4 & 3 & 8 & \frac{23}{24} & \frac{7}{6} \\
\Sp(n) & 1 & 0 & n & \frac{n}{24} & \frac{n}{12} \\
E_6 & 6 & 5 & 16 & \frac{41}{24} & \frac{13}{6} \\
E_7 & 8 & 7 & 24 & \frac{59}{24} & \frac{19}{6} \\
E_8 & 12 & 11 & 40 & \frac{95}{24} & \frac{31}{6} \\
\hline
F_4 & 5 & 4 & 12 & \frac{4}{3} & \frac{5}{3}  \\
G_2 & \frac{10}{3} & \frac{7}{3} & \frac{16}{3} & \frac{17}{24} & \frac{5}{6} 
\end{array}
\]
\caption{The cases compatible with conformal symmetry in four dimensions. For $\SU(2)$ the parameter $x$ can not be fixed by our method. Those except $F_4$ and $G_2$ are known to exist. The $F_4$ case suffers from the mismatch of the global anomaly.
\label{4ddata}}
\end{table}

Second  we consider 4d \Nequals{2} theories.
We find that the anomaly polynomials on the Higgs branch can be consistently matched only for
\begin{equation}
\SU(2),\quad
\SU(3),\quad 
\SO(8),\quad
\Sp(n),\quad
 E_{6,7,8},\quad
 F_4,\quad \text{and}\quad G_2. 
\end{equation}
The data is summarized in Table~\ref{4ddata}, using the standard notations.
The list of the groups we found here is equal to the list of group compatible with the one-instanton moduli space as Higgs branches, determined using the conformal bootstrap in \cite{Beem:2013sza,Lemos:2015orc}.\footnote{Note that in table 4 of \cite{Beem:2013sza} $\Sp(n)$ is missing. This is because the authors of \cite{Beem:2013sza} assumed that the theory is interacting in constructing their table 4.
The authors thank L. Rastelli for the clarifications. }
Note however that the $F_4$ case does not pass the global anomaly matching test, as will be detailed in Sec.~\ref{sec:4dglobal}.

\subsection{Two-dimensional  theories}
Third, we consider 2d \Nequals{(4,0)} theories.
In two dimensions, the scalars always fluctuate all over the moduli space, and the continuous symmetry never breaks.
Therefore, it is not technically correct to speak of the theory at the origin of the moduli space and compare the anomaly computed at the generic point.
Rather, what we do is to match the anomaly polynomial as calculated using a semi-classical analysis at the generic point using the unbroken symmetry at that point, with the anomaly polynomial written in terms of the full symmetry. 

We find that the anomaly polynomials on the Higgs branch can consistently be matched only for \begin{equation}
\SU(2),\quad
\SU(3),\quad 
\SO(8),\quad
\Sp(n),\quad
 E_{6,7,8},\quad
 F_4,\quad \text{and}\quad G_2.
\end{equation}
The data is summarized in Table~\ref{2ddata}, where $n_v$, $d_H$, $k_G$ are the coefficients in the anomaly polynomial expanded as follows: \begin{equation}
I^{\text{full}}_4 = -n_v c_2(R) + d_H c_2(I) + \frac{2d_H - n_f}{24} p_1(T) + \frac{k_G}4 \Tr(F^2_G)
\end{equation} where $\SU(2)_R$ and $\SU(2)_I$ are the R-symmetries.
Note that there is no global anomaly test in 2d.

\begin{table}[h]
\[
\begin{array}{c||c||c|c|c}
G & n& n_v & d_H & k_G   \\
\hline
\SU(2) &  & x-1 & 1 & x \\
\Sp(n) &  & 0 & n & 1 \\
\hline
\SU(3) & 3 & 2 & 2 & 3 \\
\SO(8) & 4 & 3 & 5 & 4 \\
F_4 & 5 & 4 & 8 & 5 \\
E_6 & 6 & 5 & 11 & 6 \\
E_7 & 8 & 7 & 17 & 8 \\
E_8 & 12 & 11 & 29 & 12  \\
\hline
G_2 & & \frac73 & 3 & \frac{10}3  
\end{array}
\]
\caption{The cases without Fermi multiplets in two dimensions. We explicitly show the value of self-Dirac-Zwazinger paring as $n$ when the theory is realized on a single string in minimal 6d \Nequals{(1,0)} theories.
\label{2ddata}}
\end{table}

\begin{itemize}
\item The $\Sp(n)$ case gives us the anomaly of $n$ free hypermultiplets gauged by $\bZ_2$.
\item  The $\SU(3)$, $\SO(8)$, $E_{6,7,8}$, $F_4$ cases reproduce the anomaly on a single string in 6d minimal gauge theories \cite{Kim:2016foj,Shimizu:2016lbw}. 
\item For the $\SU(2)$ case, we cannot completely determine the anomalies. 
\item For the $G_2$ case, we do not know any example of 2d theories with these values of anomalies.
\end{itemize}

In two dimensions, we can slightly generalize the situation by allowing the massless Fermi multiplets on the Higgs branch.  The inclusion of Fermi multiplets  opens the possibility of matching the anomaly  even for larger $\SU(n)$ and $\SO(n)$ groups. We analyze several examples with relatively simple Fermi multiplet spectrum and reproduce the anomaly of a single string in 6d non-anomalous gauge theory with various matter hypermultiplets, 
as we will show in detail in Sec.~\ref{sec:2d}.

\subsection{Cases with pure gauge anomalies and/or gauge-R anomalies}
In two, four and six dimensions, we find that for larger $\SU(n)$ and $\SO(n)$ groups, 
we cannot consistently match the anomaly polynomial.
Still, if we ignore the matching of the terms associated to $\UU(1)_F$ and $\SU(2)_F$ (which are subgroups of unbroken flavor symmetries as given in Table~\ref{basicdata}),
we find that our method somehow reproduces the values of anomalies which one would naively associate to the ADHM gauge theories realizing the one-instanton moduli spaces of these groups. 

In four dimension, these theories are infrared free, and and have a mixed gauge-gauge-R anomaly.
Moreover, for $\SO(\text{odd})$, the gauge group has the global anomaly.
In two and six dimensions, these theories have a gauge anomaly. 

We do not understand why the anomaly matching partially works for these cases.
It seems that the anomalies involving the gauge fields plays the role. 
We hope to come back to study this case further.

\newpage

\section{Six-dimensional theories}
In this section we perform the analysis for 6d \Nequals{(1,0)} theories. 

\subsection{$G$ is one of the exceptionals} First, let us specialize to the exceptional groups. Since there is no independent quartic Casimir for exceptional groups, the anomaly polynomial at the origin can be written as
\begin{align}
I^{\text{origin}}_8 &= \alpha c_2(R)^2 + \beta c_2(R) p_1(T) + \gamma p_1(T)^2 + \delta p_2(T) \nonumber \\
&+  \frac14 \Tr F^2_G \biggl{(} \frac{\kappa}4 \Tr F^2_G + \lambda c_2(R) + \mu p_1(T) \biggr{)},\label{eq:unknown6d}
\end{align}
where $c_2(R)$ is the second Chern class of the $SU(2)$ R-symmetry bundle and $p_1(T), p_2(T)$ are the first and second Pontryagin classes of the tangent bundle respectively. We have also introduced the unknown coefficients $\alpha, \beta, \gamma, \delta, \kappa, \lambda, \mu$ to be determined below. On the generic point of the Higgs branch, using \eqref{eq:2ndexp} we see that the anomaly polynomial \eqref{eq:unknown6d} becomes
\begin{align}
I^{\text{generic}}_8 &= (\alpha + \kappa + \lambda )c_2(D)^2 + (\beta + \mu) c_2(D) p_1(T) + \gamma p_1(T)^2 + \delta p_2(T) \nonumber \\
&+ \frac{m}{4} \Tr F^2_{G'} \biggl{(} \frac{m \kappa}4 \Tr F^2_{G'} + (2\kappa + \lambda) c_2(D) + \mu p_1(T) \biggr{)}.\label{eq:generic6d}
\end{align}
On the other hand, the anomaly of free hypers is given as
\begin{align}
I_8^{\text{hypers}}&= \frac{1}{24}c_2(D)^2 +  \frac1{48}c_2(D)p_1(T) + \frac{7(2+d_{R'})}{11520} p_1(T)^2 - \frac{2+d_{R'}}{2880}p_2(T) \nonumber \\
&+\frac1{48} \tr_{R'} F^4_{G'} + \frac{T^{G'}(R')}{96} \Tr F^2_{G'} p_1(T).
\label{eq:hyperanomaly6d}
\end{align} 

In order to match \eqref{eq:generic6d} and \eqref{eq:hyperanomaly6d}, there should also be no independent quartic Casimir invariant for $G'$. This already excludes $G=E_6, E_7, F_4$ and the remaining possibilities are $G=G_2, E_8$. 

\paragraph{When $G=E_8$:}  Since $\tr_{\mathbf{56}} F^4_{E_7} = \frac32 (\Tr F^2_{E_7})^2$ and $T^{E_7}(\mathbf{56})=6$, 
the anomaly \eqref{eq:hyperanomaly6d} becomes
\begin{align}
I_8^{\text{hyper}}&= \frac{1}{24}c_2(D)^2 +  \frac1{48}c_2(D)p_1(T) + \frac{203}{5760} p_1(T)^2 - \frac{29}{1440}p_2(T)+ \nonumber \\
 &+\frac1{32} \biggl{(}\Tr F^2_{G'}\biggr{)}^2+ \frac{1}{16} \Tr F^2_{G'} p_1(T). \label{eq:hyperanomaly6d"}
\end{align}
Comparing \eqref{eq:generic6d} and \eqref{eq:hyperanomaly6d"}, we can solve as
\begin{equation}
\alpha = \frac{13}{24}, \; \beta = -\frac{11}{48}, \; \gamma = \frac{203}{5760}, \; \delta = -\frac{29}{1440}, \; \kappa = \frac12, \; \lambda = -1, \; \mu = \frac14
\end{equation}
which coincides with the anomaly of rank-1 E-string theory determined in \cite{Ohmori:2014pca}.

\paragraph{When $G=G_2$:} Using $\tr_{\mathbf{4}} F^4_{\SU(2)} = \frac{41}{4} (\Tr F^2_{\SU(2)})^2$ and $T^{\SU(2)}(\mathbf{4}) = 5$, the anomaly \eqref{eq:hyperanomaly6d} becomes
\begin{align}
I_8^{\text{hypers}}&= \frac{1}{24}c_2(D)^2 +  \frac1{48}c_2(D)p_1(T) + \frac{7}{1920} p_1(T)^2 - \frac{1}{480}p_2(T)+ \nonumber \\
 &+\frac{41}{192} \biggl{(}\Tr F^2_{\SU(2)}\biggr{)}^2+ \frac{5}{96} \Tr F^2_{\SU(2)} p_1(T). \label{eq:hyperanomaly6d''}
\end{align}
Comparing \eqref{eq:generic6d} and \eqref{eq:hyperanomaly6d''}, we can solve finding
\begin{equation}
\alpha = \frac{91}{216}, \; \beta = -\frac{7}{144}, \; \gamma = \frac{7}{1920}, \; \delta = - \frac{1}{480}, \; \kappa = \frac{41}{108}, \; \lambda = -\frac{41}{54}, \; \mu = \frac{5}{72} .
\end{equation}

\subsection{$G$ is $\SU(2)$}
Let us consider the $\SU(2)$ case, which is quite exceptional. In this case, the anomaly of the hypermultiplet is just
\begin{equation}
I^{\text{hypers}}_8 = \frac{c_2(D)^2}{24} + \frac{c_2(D)p_1(T)}{48} + \frac{7p_1(T)^2 - 4p_2(T)}{5760}. \label{APFHSU26}
\end{equation}

The anomaly polynomial of the SCFT is still of the form (\ref{eq:unknown6d}) since there is no independent quartic Casimir. On a generic point of the Higgs branch $\SU(2)_G$ and $\SU(2)_R$ are identified which is implemented by \eqref{eq:2ndSU2}. Matching the resulting anomaly polynomial with (\ref{APFHSU26}), we find:
\bea
& & \alpha + \kappa + \lambda = \frac{1}{24} \label{SU2AP6}, \nonumber \\ 
& & \beta + \mu = \frac{1}{48}, \nonumber \\
& & \gamma = \frac{7}{5760}, \delta = \frac{1}{1440}.\label{6dsu2}
\eea

In this case we cannot determine the anomaly polynomial completely. The known 6d theory that have the Higgs branch $M_{\SU(2)}$ is the $\OO(1) \times \SU(2)$ free hyper. The anomaly polynomial of this theory is consistent with (\ref{SU2AP6}) with $\alpha = \beta = \lambda =0$.





\subsection{$G$ is $\SU(3)$} 

The case of $\SU(3)$ is also exceptional since the fourth Casimir of $\SU(3)$ is zero and we can take the SCFT anomaly polynomial to be of the form (\ref{eq:unknown6d}). Substituting the decomposition \eqref{eq:2ndSU3} to (\ref{eq:unknown6d}), we obtain
\begin{align}
I^{\text{generic}}_8 &= (\alpha + \kappa + \lambda) c_2(D)^2 + (\beta+\mu) c_2(D) p_1(T) + \gamma p_1(T)^2 + \delta p_2(T) \nonumber \\
&-  3 c_1 (\UU(1)_F)^2 \biggl{(} (\lambda+2\kappa) c_2(D) + \mu p_1(T) - 3 \kappa c_1 (\UU(1)_F)^2 \biggr{)} .
\end{align}

In turn the anomaly polynomial of the free hypers is given by:
\begin{multline}
I_8^{\text{hypers}}= \frac{c_2(D)^2}{24} + \frac{c_2(D)p_1(T)}{48}\\
 - \frac{3 c_1 (\UU(1)_F)^2 p_1(T)}{16} + \frac{27 c_1 (\UU(1)_F)^4}{8} + 2\frac{7p_1(T)^2 - 4p_2(T)}{5760}.
\end{multline}

Comparing the two we find a unique solution:
\bea
& & \alpha=\frac{5}{12}, \beta= - \frac{1}{24}, \nonumber \\
& & \gamma=\frac{7}{2880}, \delta= - \frac{1}{720}, \nonumber \\
& & \kappa=\frac{3}{8}, \lambda= - \frac{3}{4}, \mu= \frac{1}{16}. \label{6dsu3}
\eea

To our knowledge, a 6d SCFT with this Higgs branch is not known.

\subsection{$G$ is of type $\Sp$} 
In this case, the anomaly of the free hypers is given by
\begin{equation}
I^{\text{hypers}}_8 = \frac1{48}\biggl{(}\tr_{\text{fund}} F^4_{\Sp(n-1)} + 2 c_2(D)^2 \biggr{)} + \frac{(2c_2(D) + \tr_{\text{fund}} F^2_{\Sp(n-1)})p_1(T)}{96} + n\frac{7p_1(T)^2 - 4p_2(T)}{5760}.\label{eq:Sphypers}
\end{equation}

Since the purely gravitational part of the anomaly can be reproduced from that of the free hypers, we focus on the R-symmetry and the flavor symmetry part written as
\begin{equation}
I^{\text{origin}}_8 = \alpha c_2(R)^2 + \beta c_2(R)p_1(T) + x \tr_{\text{fund}} F^4_{\Sp(n)} + y (\tr_{\text{fund}} F^2_{\Sp(n)})^2 + \tr_{\text{fund}}F^2_{\Sp(n)} \biggl{(}\kappa c_2(R) + \lambda p_1(T) \biggr{)}. \label{AnomPolorg6d}
\end{equation}
Decomposing the characteristic classes for $\Sp(n)$ to those for $\Sp(n-1)$ using \eqref{eq:2ndSp} and \eqref{eq:4thSp}, we find that the anomaly becomes:
\begin{multline}
I^{\text{generic}}_8 = (\alpha+2x + 4y + 2\kappa) c_2(D)^2 + (\beta+2\lambda) c_2(D)p_1(T) \\
+x \tr_{\text{fund}} F^4_{\Sp(n-1)} + 4y c_2(D) \tr_{\text{fund}}F^2_{\Sp(n-1)} \\
+ y \biggl{(}\tr_{\text{fund}} F^2_{\Sp(n-1)}\biggr{)}^2 + \tr_{\text{fund}} F^2_{\Sp(n-1)} \biggl{(}\kappa c_2(D) + \lambda p_1(T)\biggr{)}.\label{eq:Spgeneric}
\end{multline}
Comparing \eqref{eq:Sphypers} and \eqref{eq:Spgeneric}, we find 
\begin{equation}
\alpha =0, \; \beta = 0, \; x= \frac1{48}, \; y=0, \; \kappa =0, \; \lambda = \frac1{96},
\end{equation}
which coincides with the anomaly of $\OO(1) \times \Sp(n)$ half-hyper when we include the purely gravitational part. This SCFT is the ADHM gauge theory for $\Sp(n)$.

\subsection{$G$ is of type $\SO$} In this case, the anomaly of the hypermultiplet is given by
\begin{align}
I^{\text{hypers}}_8 &= \frac{(n-4)\tr_{\text{fund}} F^4_{F} +6 \tr_{\text{fund}} F^2_F \tr_{\text{fund}} F^2_{\SO(n-4)} + 2\tr_{\text{fund}} F^4_{\SO(n-4)}+ 2c_2(D)^2}{48}  \nonumber \\
& +\frac{((n-4)\tr_{\text{fund}}F^2_{F} + 2\tr_{\text{fund}} F^2_{\SO(n-4)}+2 c_2(D) ) p_1(T)}{96} + (n-3)\frac{7p_1(T)^2 - 4p_2(T)}{5760}.\label{eq:SOhypers}
\end{align}
Since the purely gravitational part reproduces the anomaly at the origin, we concentrate on the part involving the R-symmetry and the flavor symmetry. We write the anomaly at the origin as
\begin{equation}
I^{\text{origin}}_8 = \alpha c_2(R)^2 + \beta c_2(R) p_1(T) + x \tr_{\text{fund}} F^4_{\SO(n)} + y (\tr_{\text{fund}} F^2_{\SO(n)})^2 + \tr_{\text{fund}} F^2_{\SO(n)} (\kappa c_2(R) + \lambda p_1(T)).\label{eq:SOorigin}
\end{equation} 
We can use equations \eqref{eq:2ndSO} and \eqref{eq:4thSO} to get:
 \bea
I^{\text{generic}}_8 & = & (\alpha+4x+16y+4\kappa) c_2(R)^2 + (\beta+4\lambda) c_2(D) p_1(T) + (x+4y) (\tr_{\text{fund}} F^2_F)^2 \\ \nonumber & + & (12x+16y+2\kappa) c_2(D) \tr_{\text{fund}} F^2_F + x \tr_{\text{fund}} F^4_{\SO(n-4)} + (16y+\kappa) c_2(D) \tr_{\text{fund}} F^2_{\SO(n-4)} \\ \nonumber & + & y (\tr_{\text{fund}} F^2_{\SO(n-4)})^2 + 4y \tr_{\text{fund}} F^2_F \tr_{\text{fund}} F^2_{\SO(n-4)} + \lambda p_1(T) (\tr_{\text{fund}} F^2_{\SO(n-4)} + 2 \tr_{\text{fund}} F^2_{F}).\label{eq:SOgeneric}
\eea
What we have to do is to match \eqref{eq:SOhypers} and \eqref{eq:SOgeneric} and solve for $\alpha, \beta, x, y, \kappa, \lambda$. We see that the $\SU(2)_F$ independent terms can be matched by setting $\alpha = -\frac{1}{8}, \beta = - \frac{1}{16}, x = \frac{1}{24}, \lambda = \frac{1}{48}, y=\kappa=0$. These are the values one get for an $\SU(2)$ gauge theory with $n$ half-hypermultiplets though it is anomalous in 6d. However it is not possible to much the remaining $\SU(2)_F$ dependent terms so there is no solution in this case.

\subsection{$G$ is of type $\SU$} We only need to consider $n\geq4$. Then, the anomaly of the hypermultiplets is given by
\bea
I^{\text{hypers}}_8 & = & \frac{c_2(D)^2}{24} + \frac{c_2(D)p_1(T)}{48} - \frac{n^2 (n-2) c_1 (\UU(1)_F)^2 p_1(T)}{48} + \frac{\tr_{\text{fund}} F^2_{\SU(n-2)} p_1(T)}{48} \nonumber \\ & - & \frac{n^2 c_1 (\UU(1)_F)^2 \tr_{\text{fund}} F^2_{\SU(n-2)}}{4} + \frac{n^4 (n-2) c_1 (\UU(1)_F)^4}{24} + \frac{\tr_{\text{fund}} F^4_{\SU(n-2)}}{24} \nonumber \\  & - & \frac{n c_1 (\UU(1)_F) \tr_{\text{fund}} F^3_{\SU(n-2)}}{6} +  (n-1)\frac{7p_1(T)^2 - 4p_2(T)}{5760}. \label{AnomPolhyp6dSU}
\eea

We take the flavor and R-symmetry part of the anomaly to be given by (\ref{AnomPolorg6d}) with the replacement $\Sp(n) \to \SU(n)$. Decomposing the characteristic classes of $\SU(n)$ into their $\UU(1)_F \times \SU(n-2)$ counterparts using \eqref{eq:2ndSU} and \eqref{eq:4thSU}, we find:
\bea
I^{\text{generic}}_8 & = & (\alpha + 2 x + 4 y + 2 \kappa) c_2(D)^2 + (\beta + 2 \lambda) c_2(D)p_1(T) + x \tr_{\text{fund}} F^4_{\SU(n-2)} + y (\tr_{\text{fund}} F^2_{\SU(n-2)})^2 \nonumber \\ & - & 8 x c_1 (\UU(1)_F) \tr_{\text{fund}} F^3_{\SU(n-2)} - 4 (6 x + n(n-2)y) c_1 (\UU(1)_F)^2 \tr_{\text{fund}} F^2_{\SU(n-2)} \nonumber \\ & - & 2 (n-2) (n \kappa + 4 y n + 6 (n-2) x ) c_1 (\UU(1)_F)^2 c_2(D) + 2 n (n-2) (x (n^2 - 6n + 12) \nonumber \\ & + & 2 y n (n-2)) c_1 (\UU(1)_F)^4 + (\kappa + 4y) c_2(D) \tr_{\text{fund}} F^2_{\SU(n-2)} \nonumber \\ & + & \lambda \tr_{\text{fund}} F^2_{\SU(n-2)} p_1(T) -4 n (n-2) \lambda c_1 (\UU(1)_F)^2 p_1(T). \label{AnomPolgen6dSU}
\eea

Matching equations (\ref{AnomPolhyp6dSU}) and (\ref{AnomPolgen6dSU}) we see that the $\UU(1)_F$ independent terms can be matched by setting $\alpha = - x = -\frac{1}{24}, \beta = - \lambda = - \frac{1}{48}, y=\kappa=0$. These are the values one get for a $\UU(1)$ gauge theory with $n$ hypermultiplets though it is anomalous in 6d. The $\UU(1)_F$ dependent terms only match if $n=2$ for which this analysis does not apply. Therefore we conclude that there is no solution in this case.


\subsection{Global anomalies} 
\label{sec:6dglobal}
Finally we consider anomalies under large gauge transformations. These exist only for groups with $\pi_6(G)\neq 0$ which are only $\SU(2), \SU(3)$ and $G_2$ for which $\pi_6(\SU(2))= \bZ_{12}, \pi_6(\SU(3))= \bZ_{6}$ and $\pi_6(G_2)= \bZ_{3}$. These anomalies are mapped to one another under the embedding of $\SU(2) \rightarrow \SU(3)\rightarrow G_2$. When embedded in groups with an independent fourth Casimir, the global anomaly can match the standard square anomaly. 

A hyper in the $\bold{7}$ of $G_2$, one in the $\bold{3}$ of $\SU(3)$, and a half-hyper in the $\bold{2}$ of $\SU(2)$ both contribute to the anomaly as the generator of $\pi_6(G)$ for their respective groups\cite{Bershadsky:1997sb}. Under the above mapping the $\bold{7}$ of $G_2$ goes to the $\bold{3} + \bar{\bold{3}}$ of $\SU(3)$ and further to the $2 \times\bold{2} + \text{singlets}$ of $\SU(2)$. Therefore the anomaly is consistently mapped across the groups. 

The only non-excluded cases where the anomaly might be relevant are $\SU(2), \SU(3)$ and $G_2$. For $\SU(2)$ and $\SU(3)$ the anomaly doesn't exist on the Higgs branch which implies that the anomaly vanishes in the SCFT. The situation for $G_2$ is more involved as it is broken to $\SU(2)$ on the Higgs branch where both groups have the discrete anomaly. 

Let's consider the $\bold{7}$ of $G_2$. Under the $\SU(2)_1 \times \SU(2)_2$ subgroup of $G_2$, it decomposes as: $\bold{7}\rightarrow (\bold{2},\bold{2}) \oplus (\bold{1},\bold{3})$. As the anomaly must be preserved, and using the fact that the $\bold{3}$ of $\SU(2)$ contribute to the anomaly like $8$ half-hyper doublets\cite{Bershadsky:1997sb}, we determine that $\SU(2)_1$ has the same anomaly as $G_2$ while $\SU(2)_2$ is non-anomalous. Therefore $\SU(2)_2$, which is the remaining global symmetry on a generic point on the Higgs branch, must be non-anomalous.

However on a generic point on the Higgs branch we have an half-hyper in the $\bold{4}$ of $\SU(2)_2$ which does contribute to the anomaly. This can be readily seen by decomposing the $\bold{14}$ of $G_2$ under the $\SU(2)_1 \times \SU(2)_2$ subgroup. Thus it is apparent that we cannot match the $\SU(2)$ anomaly with the anomaly of $G_2$.

\section{Four-dimensional theories}

In this section we implement the strategy given in Section~2 for 4d \Nequals2 theories.
We perform the analysis assuming that the theories in question are superconformal.
The analysis is slightly different depending on whether $G$ is a group of type $\SU$, $\SO$ or $\Sp$ and the exceptional groups. We next discuss each in turn.

\subsection{$G$ is of type $\Sp$ or one of the exceptionals}

When the group is of type $\Sp$ or the exceptionals then the symmetry $G'$ is a simple group. We take the anomaly polynomial of the theory to be:
\be
I^{\text{origin}}_6 =-\frac{d_H}{3} c_1(R)^3 + \frac{d_H}{12} p_1(T) c_1(R) - n_v c_1(R) c_2(R) + \frac{k_G}{4} c_1(R) \Tr F^2_G \label{eq:AnomPol4d}
\ee
where $c_1(R)$ is the first Chern class of the $\mathcal{N}$$=2$ $U(1)$ R-symmetry bundle. The form of the anomaly polynomial is dictated by \Nequals2 SUSY\cite{KT}. The constants $d_H$ and $n_v$ are related to the central charges $a, c$ through: $d_H = 24(c-a), n_v = 4(2a-c)$. The constant $k_G$ is the central charge of the flavor symmetry $G$.   

The anomaly polynomial of the free hypers is:
\be
I^\text{hypers}_6=-\frac{2+d_{R'}}{6} c_1(R)^3 + \frac{2+d_{R'}}{24} p_1(T) c_1(R) + c_1(R) c_2(D) + \frac{T^{G'}(R')}{2} c_1(R)  \Tr F^2_{G'}. \label{eq:Anom4dhypers}
\ee

Next we need to decompose the $G$-characteristic classes to the $SU(2)_D \times G'$ ones, where the relation is given in \eqref{eq:2ndexp}. By matching \eqref{eq:AnomPol4d} and \eqref{eq:Anom4dhypers} we find:
\begin{align}
d_H & =  \frac{d_{R'} + 2}{2}, &
n_v & =  \frac{2 T^{G'}(R')}{m} - 1 , &
k_G & = \frac{2 T^{G'}(R')}{m} .
\end{align}
The values of the Dynkin index are in table~\ref{index}. The complete results are summarized in table \ref{4ddata}.

For $\Sp(n)$ these are just the values of $n$ free hypers. The SCFT consisting of an $\mathrm{O}(1) \times \Sp(n)$ half-hyper indeed has this space as its Higgs branch. 

\subsection{$G$ is of type $\SO$}

In this case the group $G'$ is $SU(2)_F \times SO(n-4)$ which is a semi-simple group. We again take (\ref{eq:AnomPol4d}) as the anomaly polynomial of the SCFT and decompose the $\SO(n)$ characteristic classes by \eqref{eq:2ndSO}, but now the half-hypers contribute:
 \begin{multline}
I^\text{hypers}_6=-\frac{N-3}{3} c_1(R)^3 + \frac{n-3}{12} p_1(T) c_1(R) + c_1(R) c_2(D) \\
 + c_1(R) \Tr F^2_{\SO(n-4)} + \frac{(n-4)}{4} c_1(R) \Tr F^2_{\SU(2)_F} 
\end{multline}


Next we can proceed to match corresponding terms. Ignoring $\SU(2)_F$ terms we found that: 
\begin{equation}
d_H = n-3, \quad n_v = 3, \quad k_{\SO(n)} = 4
\end{equation}
Interestingly these are exactly the values for an $\SU(2)$ gauge theory with $n$ half-hypers which classically has this space as its Higgs branch. This is despite the fact that this theory has a global gauge anomaly for $n$ odd and even for $n$ even is not an SCFT unless $n=8$.

Finally we need to match the last $\SU(2)_F$ dependent term. This leads to the constraint $n-4 = k_{\SO(n)}$ which is only obeyed if $n=8$. 

\subsection{$G$ is of type $\SU$}

In this case the group $G'$ is $\UU(1)_F \times \SU(n-2)$. We again take (\ref{eq:AnomPol4d}) as the anomaly polynomial of the SCFT, but now the half-hypers contribute:
 \begin{multline}
I^\text{hypers}_6=-\frac{n-1}{3} c_1(R)^3 + \frac{n-1}{12} p_1(T) c_1(R) + c_1(R) c_2(D) \\
+ \frac{1}{2} c_1(R) \Tr F^2_{\SU(n-2)} - n^2 (n-2) c_1(R) c_1(\UU(1)_F)^2
\end{multline}

\paragraph{When $n>3$:} Assuming $n> 3$ and ignoring $\UU(1)_F$ terms we find that: 
\begin{equation}
d_H = n-1, \quad n_v = 1, \quad  k_{\SU(n)} = 2.
\end{equation}
Interestingly these are exactly the values for a $\UU(1)$ gauge theory with $n$ hypers which classically has this space as its Higgs branch. This is despite the fact that this theory is not an SCFT.
Indeed, to match the last $\UU(1)_F$ dependent term, we need the constraint $n^2 (n-2) = 2 n (n-2)$ which has the solution $n=2$. This is incompatible with $n>3$.

\paragraph{When $n=3$:}
For $n=3$, we now only have $\UU(1)_F$ and so $\Tr(F^2_{SU(n-2)})$ vanishes. Matching terms we find:
 \begin{equation}
d_H  =  2, \quad
n_v  =  2 ,\quad
k_G = 3 
\end{equation}
These are precisely the values of the AD $\SU(3)$ theory.

\paragraph{When $n=2$:}
For $\SU(2)$ we have only $\SU(2)_D$ as a remaining global symmetry and so we only get the constraints: 
\begin{equation}
d_H = 1,  \qquad n_v = k_{\SU(2)}-1.
\end{equation}
 These are obeyed for both the $\mathrm{O}(1)\times \SU(2)$ half-hyper and the $\SU(2)$ AD theory, which are the SCFTs known to have $M_{\SU(2)}$ as their Higgs branch.\footnote{Other known examples of the SCFT whose Higgs branch is $M_{\SU(2)}{=}\bC^2/\bZ^2$ include the superconformal point of $\SO(4k+2)$ SYM \cite{Argyres:2012fu}. However, these theories do not fit within the class of theories considered in this paper. At the generic point on the Higgs branch, the spectrum we obtain is not free hypermultplets but the interacting SCFT without Higgs branch, i.e. the superconformal fixed point of $\SU(2k-1)$ SYM. This can be readily seen by the class S description of the SCFTs.} Additionally it is obeyed for a $\UU(1)$ gauge theory with two charge $+1$ hypermultiplets, even though it is not an SCFT. 

\subsection{Global anomalies}
\label{sec:4dglobal}
So far we have used local anomalies to constraint properties of 4d \Nequals2 theories that have the one-instanton moduli space $M_G$ as their Higgs branch. We can put one additional constraint using anomalies under large gauge transformation of \cite{Witten:1982fp}. These exist only for groups with $\pi_4(G) \neq 0$, which are only $\Sp$ groups for which $\pi_4(\Sp(n)) = \bZ_2$.
When $\Sp$ group is embedded in $\Sp$ group, the global anomaly should match the global anomaly. 
When $\Sp$ group is embedded in $\SU$ group, the global anomaly can match the standard triangle anomaly \cite{Elitzur:1984kr}.

In our case this implies a non-trivial constraint only for $G=\Sp(n), F_4, G_2$. In the first case, $G=\Sp(n)$, the unbroken group on the Higgs branch is $\Sp(n-1)$, and as the matter content is a single fundamental half-hyper, it suffers from this anomaly. This can be accommodated in the SCFT if the original $\Sp(n)$ also has the same anomaly. This again agrees with the expectation from the ADHM construction.

Both $G_2$ and $F_4$ cannot have an anomaly. However, they break on the Higgs branch to groups that can, $\SU(2)$ for $G_2$ and $\Sp(3)$ for $F_4$. Therefore for these to be possible the anomaly must vanish on the Higgs branch. This is true for $G_2$ as the $\bold{4}$ of $SU(2)$ does not contribute to the anomaly. However, this is not true for $F_4$ as the $\bold{14}'$ of $Sp(3)$ does contribute to the anomaly. Thus this appears to exclude $F_4$ but leaves the possibility for $G_2$, under the very reasonable assumption that no massive degrees of freedom can provide the global anomaly for $\pi_4(\Sp(n))$.\footnote{The authors thank Kazuya Yonekura for the discussions on this point.}

\section{Two-dimensional theories}
\label{sec:2d}
In this section we analyze the 2d \Nequals{(0,4)} theories. We denote the R-symmetry as $\SU(2)_R \times \SU(2)_I$ and the general form of the anomaly polynomial is written as
\begin{equation}
I^{\text{full}}_4 = -n_v c_2(R) + d_H c_2(I) + \frac{2d_H - n_f}{24} p_1(T) + \frac{k_G}4 \Tr(F^2_G),\label{eq:unknown2d}
\end{equation}
where $n_V$,$d_H$, $n_F$ and $k_G$ are the unknown coefficients determined below. Note that the $\SU(2)_I$ and the gravitational part of the anomaly can be matched directly on the Higgs branch. We also note that there are no global gauge anomalies in 2d since $\pi_2(G)=0$ for all Lie groups. 

In 2d, we can consider the slightly generalized situation: we can also have Fermi multiplets in addition to hypermultiplets on a generic point of the Higgs branch. Fermi multiplet consists of a single left-moving Weyl fermion transforming some representation $R_F$ under $G$.\footnote{Some references (e.g. \cite{Tong:2014yna}) define a pair of left-moving Weyl fermions transforming in conjugate representations as a 2d \Nequals{(0,4)} Fermi multiplet. Here we choose not to use such a definition.} In this section, we also examine how the anomaly matching changes when we allow the Fermi multiplets as the massless spectrum.\footnote{In this note, we only consider Fermi multiplets transforming non-trivially under $G'$. The effect of neutral Fermi multiplets is to change the value of the gravitational anomaly.}

\subsection{$G$ is of type $\Sp$ or one of the exceptionals} In this case, the unbroken subgroup $G'$ is simple. If we denote the representations of Fermi multiplets under $G'$ as $\sum_m \mathbf{N_m}$, then the anomaly polynomial of free multiplets is given as
 \begin{equation}
I_4^{\text{free}}=  c_2(D) + \frac{2+d_{R'}}2  c_2(I) + \frac{2+d_{R'}-\sum_m N_m}{24} p_1(T) + \frac{2T^{G'}(R')-2\sum_mT^{G'}(\mathbf{N_m})}4 \Tr(F_{G'}^2). \label{eq:hyperanomaly2d}
\end{equation} 
On the other hand, by using \eqref{eq:2ndexp} and \eqref{eq:2ndSp}, the anomaly \eqref{eq:unknown2d} becomes
\begin{equation}
I^{\text{generic}}_4 = (k_G - n_v) c_2(D) + d_H c_2(I) + \frac{2d_H - n_f}{24} p_1(T) + \frac{mk_G}4 \Tr(F^2_{G'}),\label{eq:generic2d}
\end{equation}
where $m$ is $3$ for $G_2$ and $1$ for other cases.

\paragraph{Without Fermi multiplets:} If we assume that there are no Fermi multiplets, the anomalies \eqref{eq:hyperanomaly2d} and \eqref{eq:generic2d} can be matched by the data summarized in table \ref{2ddata}. 

The cases with $G=E_8, E_7, E_6, F_4$ reproduce the anomaly on a single self-dual string\footnote{We have subtracted the anomaly of the center-of-mass mode from the result presented in \cite{Shimizu:2016lbw}.} in minimal 6d \Nequals{(1,0)} theories for $n=12,8,7,5$:
\begin{equation}
I_4^{\text{string}} = -(n-1)c_2(R) + (3n -7) c_2(I) + \frac{3n-7}{12}p_1(T) + \frac{n}4 \Tr F^2_G. \label{eq:inflow2d}
\end{equation}
The case with $G=\Sp(n)$ reproduces the anomaly of $O(1) \times \Sp(n)$ half-hypers as in 4d and 6d. To the best of our knowledge, we do not know an example of 2d \Nequals{(0,4)} SCFT with Higgs branch $M_{G_2}$ and no Fermi multiplets.

\paragraph{With Fermi multiplets:} Next we consider the cases with Fermi multiplets on the Higgs branch. 

As examples, let us consider $n_f$ fundamental Fermi multiplets of $G'$. 
For the $G=E_7$, the anomaly is given by
\begin{equation}
I^{\text{full}}_4 = -( 7- n_f)c_2(R) +17c_2(I) + \frac{17-3n_f}{12} p_1(T) + \frac{8-n_f}4 \Tr (F^2_{E_7})+\frac14 \Tr (F^2_{\SO(n_f)}).  
\end{equation}
where we included the $\SO(n_f)$ symmetry acting on Fermi multiplets. This anomaly precisely agrees  with that of a single string in 6d $E_7$ gauge theory with $n_f /2$ hypermultplets. Similarly, $G=E_6, F_4$ cases reproduce the anomaly of a single string in 6d $G=E_6, F_4$ gauge theory with $n_f$ fundamental hypermultiplets.

Finally, we consider $G=G_2$. The anomaly can be matched by
\begin{equation}
I^{\text{full}}_4 = -\frac{7-n_f}{3}c_2(R) + 3c_2(I) + \frac{3-n_f}{12}p_1(T) + \frac{10-n_f}{12} \Tr (F^2_{G_2}) + \frac14 \Tr (F^2_{\SU(n_f)}),\label{eq:2dG2anomaly}
\end{equation}
where we included the $\SU(n_f)$ flavor symmetry acting on the Fermi multiplets. For $n_f = 1,4,7$, \eqref{eq:2dG2anomaly} reproduces the anomaly of a string in the 6d $G_2$ gauge theory with $n_f=1,4,7$ fundamental hypermultiplets. 

\subsection{$G$ is of type $\SO$} In this case, the unbroken group is $\SU(2)_F \times \SO(n-4)$. If we denote the representation of the Fermi multiplets by $\sum_m (\mathbf{n_m}, \mathbf{N_m})$, the anomaly of the free multiplets is given by
\begin{align}
I^{\text{free}}_4 &= c_2(D) + (n-3) c_2(I) + \frac{2n-6-\sum_m n_m N_m}{24} p_1(T)  \nonumber \\
&+ \frac{n-4-2\sum_m N_m T^{\SU(2)_F}(\mathbf{n_m})}4 \Tr (F^2_F) + \frac{2-\sum_m n_m T^{\SO(n-4)}(\mathbf{N_m})}2 \Tr (F^2_{\SO(n-4)}). \label{eq:2dSOhypers} 
\end{align}

On the other hand, by using \eqref{eq:2ndSO}, the anomaly \eqref{eq:unknown2d} becomes
\begin{equation}
I^{\text{generic}}_4 = (k_G - n_v)c_2(D) + d_H c_2(I) + \frac{2d_H - n_F}{24} p_1(T) + \frac{k_G}4 \Tr (F^2_F) +\frac{k_G}4 \Tr (F^2_{\SO(n-4)}),   \label{eq:2dSOgeneric}
\end{equation}

\paragraph{Without Fermi multiplets:} Comparing \eqref{eq:2dSOhypers} and \eqref{eq:2dSOgeneric} in the case of Fermi multiplets, the anomaly can be solved by
\begin{equation}
n_v =3, \qquad d_H =n-3, \qquad k_G = 4.
\end{equation}
if we ignore the $\SU(2)_F$ part. This precisely agrees with the values of the $\SU(2)$ gauge theory with $n$ half-hypers, though it is anomalous in 2d. If we include the matching of $\SU(2)_F$, the solution exists only for $G=\SO(8)$ and we obtain the anomaly of \eqref{eq:inflow2d} for $n=4$. Indeed, the worldsheet theory on a single string in minimal 6d \Nequals{(1,0)} SCFT for $n=4$ has the Higgs branch $M_{\SO(8)}$. 

\paragraph{With Fermi multiplets:} Let us consider the cases with Fermi multiplets. The matching of $\SU(2)_F$ puts a constraint
\begin{equation}
n-4-2\sum_m N_m T^{\SU(2)_F}(\mathbf{n_m}) = 4-2\sum_m n_m T^{\SO(n-4)}(\mathbf{N_m})
\end{equation}
An example of solution of these constraints is obtained by setting $1\leq m \leq n-8$, $\mathbf{N_m}=\mathbf{1}$ and $\mathbf{n_m}=\mathbf{2}$ for all $m$. The anomaly polynomial is 
\begin{equation}
I^{\text{full}}_4 = -3 c_2(R) + (n-3)c_2(I) +\frac{5}{12}p_1(T) + \Tr (F^2_{\SO(n)}) + \frac14 \Tr (F^2_{\Sp(n-8)}),
\end{equation}
where we have included the global symmetry acting on $(n-8)$ free Fermi multiplets. This is precisely the anomaly of a single string in 6d $\SO(n)$ gauge theory with $(n-8)$ fundamental hypermultiplets. 

\subsection{$G$ is of type $\SU$} If we denote the representation of the Fermi multiplets as $\oplus_{m}(\mathbf{N_m})_{\mathbf{n_m}}$ under $\SU(n-2) \times \UU(1)_F$, the anomaly of the free multiplets is 
\begin{align}
I^{\text{free}}_4 &= c_2(D) + (n-1) c_2(I) + \frac{2n-2-\sum_{m}N_m}{24} p_1(T) +\frac{2-2\sum_{m} T^{\SU(n-2)}(\mathbf{N_m})}{4} \Tr (F^2_{\SU(n-2)}) \nonumber \\
&- \biggl{(}n^2 (n-2) - \frac12 \sum_m N_m n_m^2 \biggr{)}c_1(\UU(1)_F)^2. \label{eq:2dSUhypers}
\end{align}
On the other hand, by using the decomposition \eqref{eq:2ndSU}, we have the anomaly 
\begin{align}
I^{\text{generic}}_4 &= (k_{\SU(n)} - n_v)c_2(D) + d_H c_2(I) + \frac{2d_H - \sum_{m}N_m}{24} p_1(T) + \frac{k_{\SU(n)}}{4} \Tr (F^2_{\SU(n-2)}) \nonumber \\
&- k_{\SU(n)} n(n-2) c_1(\UU(1)_F)^2.\label{eq:2dSUgeneric}
\end{align}

\paragraph{Without Fermi multiplets:} Let us first consider the case $n \geq 4$. If we ignore the $\UU(1)_F$ part, the matching between \eqref{eq:2dSUhypers} and \eqref{eq:2dSUgeneric} can be solved by
\begin{equation}
n_v =1, \qquad d_H = n-1, \qquad k_{\SU(n)}=2
\end{equation}
which precisely agrees with the values of the $\UU(1)$ gauge theory with $n$ hypermultiplets, though it is anomalous in 2d. The matching of $\UU(1)_F$ forces us to set $n=2$, which contradicts with our assumption.

When $G=\SU(2)$, the matching can be solved by 
\begin{equation}
I^{\text{full}}_4 = - (k_{\SU(2)}-1) c_2(R) + c_2(I) + \frac1{12} p_1(T) + \frac{k_{\SU(2)}}4 \Tr (F^2_{\SU(2)}), 
\end{equation}
where $k_{\SU(2)}$ is an undetermined coefficient. If we set $k_{\SU(2)}=1$, we reproduce the anomaly of the $\OO(1) \times \SU(2)$ half-hyper as in 4d and 6d.

When $G=\SU(3)$, the matching can be solved by
\begin{equation}
I^{\text{full}}_4 = - 2 c_2(R) + 2c_2(I) + \frac1{6} p_1(T) + \frac34 \Tr (F^2_{\SU(3)} ), 
\end{equation}
which coincides with the anomaly \eqref{eq:inflow2d} for $n=3$. Indeed, the worldsheet theory of a single string in minimal 6d \Nequals{(1,0)} SCFT for $n=3$ is the SCFT with Higgs branch $M_{\SU(3)}$.

\paragraph{With Fermi multiplets:} Let us consider the case with Fermi multiplets for $n \geq 4$. 
We consider two cases.
When $\mathbf{N_{m \geq 1}}=\mathbf{1}$, the matching can be solved by
\begin{equation}
n_V =1, \qquad d_H = n-1, \qquad k_{\SU(n)}=2
\end{equation}
as long as the $\UU(1)_F$ charges satisfy
\begin{equation}
2n(n-2) = n^2 (n-2) - \frac12 \sum_{m} n^2_m. \label{eq:U1const}
\end{equation}
An example of solutions of \eqref{eq:U1const} is obtained by setting $1 \leq m \leq 2n$ and $n_m = (n-2)$ for all $m$. The full anomaly is 
\begin{equation}
I_4^{\text{full}}= -c_2(R) + (n-1)c_2(I) - \frac1{12}p_1(T) + \frac24 \Tr(F^2_{\SU(n)}) + \frac14 \Tr(F^2_{\SU(2n)_F}),
\end{equation}
where we have included the contribution of the flavor symmetry $\SU(2n)_F$, acting on the Fermi multiplets of the same $\UU(1)_F$ charges. This is precisely the anomaly of a single string in 6d $\SU(n)$ gauge theory with $2n$ fundamental hypermultiplets.

When $\mathbf{N_1}=(\mathbf{n-2}), \mathbf{N_{m\geq2}=\mathbf{1}}$, the matching can be solved by
\begin{equation}
n_v =0, \qquad d_H = n-1, \qquad k_{\SU(n)}=1
\end{equation}
as long as the $\UU(1)_F$ charges satisfy 
\begin{equation}
n(n-2)= n^2 (n-2) - \frac{(n-2)n_1^2}2  - \frac12 \sum_{m\geq2} n_m^2.\label{eq:U1const2}
\end{equation} 
An example of solutions of \eqref{eq:U1const2} is obtained by setting $1\leq m \leq n+9$, $n_1 = (n-4)$ and $n_m = (n-2)$ for all $m\geq2$. The total anomaly is given by
\begin{equation}
I^{\text{full}}_4 = (n-1)c_2(I) - \frac13 p_1(T) + \frac14 \Tr (F^2_{\SU(n)}) + \frac14 \Tr (F^2_{\SU(n+8)_F}),
\end{equation}
where we have included the global symmetry $\SU(n+8)$ acting on the Fermi multiplets of the same $\UU(1)_F$ charge. This precisely agrees with the anomaly of a single string in 6d $\SU(n)$ gauge theory with $N_f =n+8$, $N_{\Lambda^2}=1$ hypermultiplets. 

\section*{Acknowledgements}
The authors would like to thank Kantaro Ohmori and Kazuya Yonekura for useful discussions.
HS is partially supported by the Programs for Leading Graduate Schools, MEXT, Japan, via the Leading Graduate Course for Frontiers of Mathematical Sciences and Physics. 
HS is also supported by JSPS Research Fellowship for Young Scientists.
The work of YT is partially supported in part by JSPS Grant-in-Aid for Scientific Research No.~25870159.
The work of YT and GZ are partially supported  by WPI Initiative, MEXT, Japan at IPMU, the University of Tokyo.

\appendix
\section{Decomposition of characteristic classes}
In this appendix, we collect the formulas relating the characteristic classes for $G$ and $G'$, used in the main body of the paper. We define the $\Tr$ by the trace in the adjoint representation, divided by the dual Coxeter number of $G$. The Dynkin index of the representation $R$ of gauge group $G$ relates the $\Tr F^2_G$ via 
\begin{equation}
\tr_{R} F^2_G = T^{G}(R) \Tr F^2_G,
\end{equation}
where $\tr_R$ is the trace in the representation $R$.
We list the values of $T^G(R)$ relevant in this paper in table~\ref{index}.

\begin{table}[h]
\[
\begin{array}{c||c|c|c|c|c|c|c|c}
G & E_7 & \SO(12) & \SU(6) & \Sp(3) & \SU(2) & \Sp(n) & \SO(n) & \SU(n) \\
\hline
R & \mathbf{56} &  \mathbf{32} & \mathbf{20} & \mathbf{14'} & \mathbf{4} & \mathbf{2n} & \mathbf{n} & \mathbf{n} \\
T^G(R) & 6 & 4  & 3 & \frac52 & 5 & \frac12 & 1 & \frac12 \\
\end{array}
\]
\caption{The values of $T^G(R)$ for various representations.
\label{index}}
\end{table}

\paragraph{When $G$ is one of the exceptionals:}
Since there are no independent quartic Casimir invariants in this case, we only have to consider $\Tr F^2_G$. The unbroken subgroup $G'$ is simple. The formula is
\begin{equation}
\Tr(F^2_{G}) = 4 c_2 (D) + m \Tr F^2_{G'},\label{eq:2ndexp}
\end{equation}
where $m=3$ for $G_2$ and $1$ for any other group.

\paragraph{When $G$ is of type $\Sp$:} The unbroken subgroup is $\Sp(n-1)$ in this case. 
The $\Tr F^2_{\Sp(n)}$ is related that of $\Sp(n-1)$ via
\begin{equation}
\Tr(F^2_{\Sp(n)}) = 4 c_2 (D) +  \Tr F^2_{\Sp(n-1)}.\label{eq:2ndSp}
\end{equation}
The $\tr_{\text{fund}} F^4_{\Sp(n)}$ is related by
\begin{equation}
\tr_{\text{fund}} F^4_{\Sp(n)} = 2c_2(D)^2 + \tr_{\text{fund}} F^4_{\Sp(n-1)}.\label{eq:4thSp}
\end{equation}

\paragraph{When $G$ is of type $\SO$:} The unbroken subgroup is $\SU(2)_F \times \SO(n-4)$ in this case. The $\Tr F^2_{\SO(n)}$ is related via
\begin{equation}
\Tr F^2_{\SO(n)} = 4 c_2 (D) + \Tr F^2_F+ \Tr F^2_{\SO(n-4)}. \label{eq:2ndSO}
\end{equation}
The $\tr_{\text{fund}} F^4_{\SO(n)}$ is related by
\be
\tr_{\text{fund}} F^4_{\SO(n)} = 4 c_2(D)^2 + 2\tr_{\text{fund}} F^4_{F} + 12 c_2(D) \tr_{\text{fund}} F^2_{\SO(n-4)} + \tr_{\text{fund}} F^4_{\SO(n-4)}. \label{eq:4thSO}
\ee

\paragraph{When $G$ is of type $\SU$:}  First, we assume $n\geq 4$. The $\Tr F^2_{\SU(n)}$ becomes
\be
\Tr F^2_{\SU(n)} = 4 c_2 (D) - 4 n (n-2) c_1(\UU(1)_F)^2 + \Tr(F^2_{\SU(n-2)}),\label{eq:2ndSU}
\ee
and $\tr_{\text{fund}}F^4_{\SU(n)}$ becomes
\begin{multline}
\tr_{\text{fund}} F^4_{\SU(n)}  = \tr_{\text{fund}} F^4_{\SU(n-2)} - 8 c_1 (\UU(1)_F) \tr_{\text{fund}} F^3_{\SU(n-2)} - 24 c_1 (\UU(1)_F)^2 \tr_{\text{fund}} F^2_{\SU(n-2)} + \\
2c_2(D)^2  -  12 (n-2)^2 c_1 (\UU(1)_F)^2 c_2(D) + 2 n (n-2)(n^2 -6 n + 12) c_1 (\UU(1)_F)^4 .\label{eq:4thSU}
\end{multline}

The cases of $\SU(2)$ and $\SU(3)$ are quite exceptional since these groups have no independent quartic Casimir and we only have to consider $\Tr F^2$. For $G=\SU(2)$, $\SU(2)_R$ is identified with the original $G$ and we simply take 
\begin{equation}
\Tr F^2_{\SU(2)}= 4c_2(D)=4c_2(R).\label{eq:2ndSU2}
\end{equation}
For the case of $\SU(3)$, we use
\be
\Tr F^2_{\SU(3)} = 4c_2 (D) - 12 c_1 (\UU(1)_F)^2.\label{eq:2ndSU3}
\ee

\bibliographystyle{ytphys}
\bibliography{higgsref}

\end{document}